\documentclass[aps,prd,onecolumn,amssymb]{revtex4}
\usepackage{graphicx,bm,color}
\usepackage[english]{babel}
\usepackage{amsmath}
\usepackage{amsfonts}

\begin{document}

\newcommand{\be}{\begin{equation}}
\newcommand{\ee}{\end{equation}}
\newcommand{\bea}{\begin{eqnarray}}
\newcommand{\eea}{\end{eqnarray}}
\newcommand{\nn}{\nonumber \\}
\newcommand{\e}{\mathrm{e}}

\title{Scalar dark energy models mimicking $\Lambda$CDM with arbitrary future
evolution}
\author{Artyom~V.~Astashenok$^{1}$, Shin'ichi Nojiri$^{2,3}$,
Sergei~D.~Odintsov$^{4,5}$\footnote{
Also at Tomsk State Pedagogical
University, Tomsk, Russia.
}, and Robert~J.~Scherrer$^6$}
\affiliation{$^1$ Baltic Federal University of I. Kant, Department of
Theoretical Physics, 236041, 14, Nevsky st., Kaliningrad, Russia \\
$^2$ Department of Physics, Nagoya University, Nagoya
464-8602, Japan \\
$^3$ Kobayashi-Maskawa Institute for the Origin of Particles and
the Universe, Nagoya University, Nagoya 464-8602, Japan \\
$^4$Instituci\`{o} Catalana de Recerca i Estudis Avan\c{c}ats (ICREA),
Barcelona, Spain \\
$^5$Institut de Ciencies de l'Espai (CSIC-IEEC),
Campus UAB, Facultat de Ciencies, Torre C5-Par-2a pl, E-08193 Bellaterra
(Barcelona), Spain \\
$^6$ Department of Physics and Astronomy, Vanderbilt University, Nashville TN
37235, United States}

\begin{abstract}
Dark energy models with various scenarios of evolution are considered from
the viewpoint of the formalism for the equation of state. It is shown that
these models
are compatible with current astronomical data.  Some of the models presented
here
evolve arbitrarily close to $\Lambda$CDM up to the present, but diverge
in the future into a number of different possible asymptotic states, including
asymptotic
de-Sitter (pseudo-rip) evolution, little rips with disintegration of bound
structures, and various forms of finite-time future
singularities. Therefore it is
impossible from observational data to determine whether the universe
will end in a future singularity or not. We demonstrate that the models under
consideration are stable for a long period of time (billions of years) before
entering a Little
Rip/Pseudo-Rip induced dissolution of bound structures or before entering a
soft finite-time future singularity. Finally, the physical consequences of Little
Rip, Type II, III and Big Crush singularities are briefly compared.

\end{abstract}

\maketitle

\section{Introduction}

The discovery of the accelerating expansion of the universe
\cite{Riess,Perlmutter} has
raised a number of difficult problems in cosmology. The cosmic
acceleration can be explained via the introduction of so-called dark energy
(for recent review, see \cite{Dark-6,Cai:2009zp}) with quite strange properties
like
negative pressure and/or negative entropy, invisibility in the early universe,
etc. According to the
latest supernovae observations, the dark energy currently accounts for
73\% of the total mass energy of the universe (see, for instance,
Ref.~\cite{Kowalski}).

The equation of state (EoS) parameter $w_\mathrm{D}$ for dark energy is
negative:
\be
w_\mathrm{D}=p_\mathrm{D}/\rho_\mathrm{D}<0\, ,
\ee
where $\rho_\mathrm{D}$ is the dark energy density and $p_\mathrm{D}$ is the
pressure. Although astrophysical observations favor the standard
$\Lambda$CDM cosmology, the uncertainties in the determination of the EoS dark
energy parameter $w$ are
still too large to define which of the three cases $w < -1$, $w = -1$, and
$w >-1$ is realized in our universe: $w=-1.04^{+0.09}_{-0.10}$
\cite{PDP,Amman}.

If $w<-1$ (phantom dark energy) \cite{Caldwell}, we are dealing with the most
interesting and least understood theoretical case.
For a phantom field, the violation of all of the four energy conditions occurs.
Although this field is unstable from the quantum field theory viewpoint
\cite{Carrol}, it could be stable in classical cosmology. Some
observations \cite{P} may be understood as the indication of the crossing
of the phantom divide in the near past or in the near future.
A very unpleasant property of phantom dark energy is the Big Rip future
singularity \cite{Caldwell,Frampton,S,BR,Nojiri},
where the scale factor becomes infinite at a finite time in the future.
A less dangerous future singularity caused by phantom or quintessence dark
energy
is the sudden (Type II) singularity \cite{Barrow} where the scale factor is
finite
at Rip time. Nevertheless, the condition $w<-1$ is not sufficient for a
singularity to occur.
Mild phantom models where $w$ asymptotically tends to $-1$ and the energy
density increases with time
or remains constant but there is no finite-time future singularity are
discussed in recent works \cite{Frampton-2,Frampton-3,Astashenok}.
The key point is that if $w$ approaches $-1$ sufficiently rapidly,
then it is possible to have a model in which the time required for
the occurrence of the singularity is infinite, i.e., the
singularity effectively does not
occur. However, if the energy density grows, the disintegration of bound
structures
eventually occurs in a way similar to the case of the Big Rip singularity.
Such phenomena, called the Little Rip and the Pseudo-Rip, destroy the bound
structures in the universe at a finite time!

The most convenient formalism to construct the (scalar or fluid)
dark energy is to use the EoS:
\be
\label{EoS-0}
p=g(\rho)\, ,
\ee
where $g$ is a function of the energy density.
The evolution of the universe then depends on the choice of the EoS.

The aim of this article is to develop a general approach for the construction
of
dark energy models which are compatible with observational data and which
provide
various scenarios of evolution.
Moreover, we are most interested in models which are stable for a long period
of time before entering the Little rip/Pseudo-Rip induced dissolution of bound
structures or before entering the mild finite-time
future singularity (with a finite scale factor at the Rip).
This question is
analyzed from the viewpoint of the dark energy EoS and
corresponding description in terms of scalar field theory. In Sec.~\ref{SecII},
the general approach to this problem is developed. 
The classification of cosmological models with different future evolutions is
presented. The confrontation with observational data is also given.
The non-singular Little Rip cosmology based on a massive scalar potential is
considered in Sec.~\ref{SecIII}. 
It is demonstrated that it may be indistinguishable from the $\Lambda$CDM
cosmology, being stable for a long time before the disintegration of bound structures. 
Although the similar models are considered in \cite{Frampton-2}, 
it is interesting to note that the simplest model with Klein-Gordon potential is 
described apparently for the first time. 
In Sec.~\ref{SecIV} the asymptotically de Sitter quintessence or phantom
cosmology is proposed. The properties of such realistic cosmologies are
similar to those of the previous section. 
In comparison, for example, with \cite{Frampton-3}, our approach is based on 
equation-of-state formalism instead of setting of the Hubble parameter as function of time. 
The next two sections are devoted to the construction of quintessence
dark energy with a Type III future singularity and phantom/quintessence
dark energy with a Type II future singularity. 
These results based on EoS formalism are novel in cosmology. 
The estimation of time remaining before Type II or Type III singularity is 
made for first time.
The comparison of the
predictions of such models with observational data demonstrates that they
may be indistinguishable the $\Lambda$CDM model up to the present. Moreover,
they may be
stable for billions of years before reaching a future singularity.
In this sense, such models represent a quite viable alternative to
$\Lambda$CDM. Sec.~\ref{SecVII} is devoted to the construction of Big Crush
scalar quintessence models and their comparison with the Little Rip or Type II
future
singularity cosmology. Specifically, some physical properties of the Big Crush
versus the Rip are discussed.  Phantom models are briefly reviewed
in Sec.~\ref{SecPhantom}.
The main qualitative result of our study is that the current observations
make it essentially
impossible to determine whether or not the universe will end in a future
singularity. 
We should note once more that the consistent application of equation-of-state formalism 
is the main feature of our work. 
Some summary and outlook are given in the Discussion section.

\section{Scalar dark energy models \label{SecII}}

In this section, we consider the construction of one-scalar dark energy
models and compare the models with observational data.
We mainly concentrate on different phantom theories whose classical
behavior is very similar to that of the $\Lambda$CDM model.

For the spatially-flat FRW universe with metric
\be
ds^{2}=dt^{2}-a^{2}(t)(dx^{2}+dy^{2}+dz^{2})\, ,
\ee
the cosmological equations, that is, the FRW equations are given by
\be
\label{Fried1}
\left(\frac{\dot a}{a}\right)^2 = \frac{\rho}{3}\, , \quad
\dot{\rho} = -3\left(\frac{\dot a}{a}\right)(\rho + p)\, ,
\ee
where $\rho$ and $p$ are the total energy density and pressure,
$a$ is the scale factor, $\dot{}=d/dt$,
and we use the natural system of units in which $8\pi G=c=1$.

It is convenient to write the dark energy equation of state (EoS)
in the following form:
\be \label{EoS}
p_\mathrm{D}=-\rho_\mathrm{D}-f(\rho_\mathrm{D})\, ,
\ee
where $f(\rho_\mathrm{D})$ is some function. The case $f(\rho_\mathrm{D})>0$
corresponds
to the EoS parameter $w<-1$ (phantom) while the case $f(\rho)<0$ corresponds to
the EoS
parameter $w>-1$. If dark energy dominates, one can neglect the contribution
of other components (matter, dark matter). Then from Eq.~(\ref{Fried1}), one
can get the following expression for time variable:
\be
\label{trho}
t = \frac{2}{\sqrt{3}}\int^{x}_{x_{0}} \frac{d x}{f(x)}\, , \quad
x\equiv\sqrt{\rho}\, .
\ee
Hereafter, it is convenient to omit the subscript $\mathrm{D}$. The
quintessence
energy density decreases with time ($x<x_{0}$), while the phantom energy
density increases ($x>x_{0}$).
Following Ref.~\cite{Nojiri}, one can find the following behavior for the
expression (\ref{trho}):
\begin{enumerate}
\item The integral (\ref{trho}) converges at $\rho\rightarrow\infty$. Therefore
a finite-time singularity occurs: the energy density becomes infinite at a
finite time $t_\mathrm{f}$. The expression for the scale factor
\be
\label{arho}
a = a_{0}\exp\left(\frac{2}{3}\int^{x}_{x_{0}} \frac{x d x}{f(x)}\right)\, ,
\ee
indicates that there are two possibilities:
\begin{enumerate}
\item\label{1a} 
The scale factor diverges at a finite time (Big Rip singularity
\cite{Caldwell,Frampton,BR}).
\item\label{1b} 
The scale factor remains finite; however, a singularity
($\rho\rightarrow\infty$) occurs.
This is a Type III singularity \cite{Nojiri-2}.
\end{enumerate}
The key difference between (\ref{1a}) 
and (\ref{1b}) 
is that for (\ref{1b}) 
the energy density grows so rapidly with time that
the scale factor does not reach an infinite value.
\item The integral (\ref{trho}) diverges at $\rho\rightarrow\infty$.
Such models are described in \cite{Frampton-2} (see also
Refs.~\cite{Astashenok,others}).
The energy density grows with time but
not rapidly enough for the emergence of the Big Rip singularity.
According to Ref.~\cite{Frampton-2}, we have a so-called ``Little Rip'':
eventually a dissolution of bound
structures occurs at a finite future time. Nevertheless, formally the future
singularity does not occur (or, rather, it is shifted to the infinite
future).
Such scenarios are possible only in the case of phantom dark energy. The next
two scenarios are possible for both phantom and quintessence dark
energy.
\item The integral (\ref{trho}) diverges at $\rho\rightarrow\rho_\mathrm{f}$.
The dark
energy density asymptotically tends to a constant value (``effective
cosmological constant''). Such asymptotically de Sitter theories represent
the natural alternative to the $\Lambda$CDM model, which also leads to
non-singular cosmology.
Nevertheless, even for a non-singular asymptotically de Sitter universe, the
possibility of a dramatic rip which may lead to disappearance of bound
structures in the universe remains possible \cite{Frampton-2,Astashenok}.
\item Another interesting case corresponds to $f(x)\rightarrow\pm \infty$ at
$x=x_\mathrm{f}$, i.e., the pressure of the dark energy becomes infinite at a
finite energy density. The second derivative of the scale factor diverges while
the first
derivative remains finite. It is interesting to investigate the
properties of dark energy with
such a (sudden or Type II) finite-time future
singularity \cite{Barrow,Nojiri-2}.
\end{enumerate}
What does this last type of singularity mean from the physical viewpoint?
As the
universe expands, the relative acceleration between two points separated
by a distance $l$ is given by $l \ddot a/a$.
If there is a particle with mass $m$ at each of the points, an observer at
one of the masses will measure an inertial force on the other mass of
\be
\label{i1}
F_\mathrm{iner}=m l \ddot a/a = m l \left( \dot H + H^2 \right)\, .
\ee
Let us assume that two particles are bounded by a constant force $F_0$. If
$F_\mathrm{iner}$ is positive and greater
than $F_0$, the two particles become unbounded. This is the rip produced by
the accelerating expansion.
Note that Eq.~(\ref{i1}) shows that the rip always occurs when either $H$
diverges or $\dot H$ diverges (assuming $\dot H > 0$). The first case
corresponds to
the Big Rip singularity, while
if $H$ is finite, but $\dot H$ diverges with $\dot H > 0$,
we have a Type II or sudden future singularity, which also leads to a rip.
Even if $H$ or $\dot H$ goes to infinity at the infinite future, the
inertial force becomes larger and larger,
and any bound object is ripped, i.e., the Little Rip cosmology emerges.
If $H$ is finite and $\dot H$ is negative and diverges, then all the structures
are crushed rather than ripped.

Eq.~(\ref{i1}) can be rewritten as
\be
\label{Fin}
F_\mathrm{iner}={m l}\left(\frac{x^{2}}{3}+\frac{f(x)}{2}\right)\, .
\ee
The crush corresponds to $f(x)\rightarrow -\infty$ at
$x\rightarrow x_\mathrm{f}<\infty$ while the case
$f(x)\rightarrow\infty$ at $x\rightarrow x_\mathrm{f}$
describes the sudden future singularity.
The equivalent description in
terms of scalar theory can be derived using the equations:
\be
\rho=\pm\dot{\phi}^{2}/2+V(\phi)\, , \quad p=\pm\dot{\phi}^{2}/2-V(\phi)\, ,
\ee
where $\phi$ is the scalar field with potential $V(\phi)$. The sign ``$-$''
before kinetic term corresponds to the phantom energy. For the scalar field and
its potential, one can derive the following expressions:
\bea
\label{phix}
&&
\phi(x)=\phi_{0}\pm\frac{2}{\sqrt{3}}\int_{x_{0}}^{x}\frac{dx}{\sqrt{|f(x)|}
}\, ,\\
\label{Vx}
&& V(x)=x^{2}+f(x)/2\, .
\eea
Combining Eqs.~(\ref{phix}) and (\ref{Vx}) gives the potential as function
of the scalar field.
For simplicity, we choose the sign ``$+$'' in Eq.~(\ref{phix}) for
phantom energy and ``$-$'' for quintessence.
For the crush and sudden future
singularity the potential of the scalar field has a pole, i.e. from the
mathematical viewpoint, these singularities are equivalent.
Note that singularities often correspond to the infinite value of
the scalar field $\phi\to \pm \infty$.
For the sudden future singularity
potential, we find $V(\phi)\rightarrow+\infty$, and for the big crush,
$V(\phi)\rightarrow -\infty$.

Confrontation of the theoretical models with observational data consists mainly
of comparison with the
distance modulus as a function of redshift from the Supernova Cosmology
Project \cite{Amanullah:2010vv}.
The distance modulus for a supernova with redshift $z=a_{0}/a-1$ is
\be
\mu(z)=\mbox{const}+5\log D(z)\, ,
\ee
where $D(z)$ is the luminosity distance. As is well-known, the
SNe data are well-fit by the $\Lambda$CDM cosmology. For such a model (which we
call the ``standard cosmology'' (SC)), we obtain
\be
\label{DLSC}
D^\mathrm{SC}_\mathrm{L}=\frac{c}{H_{0}}(1+z)\int_{0}^{z}
\left(\Omega_{m}(1+z)^{3}+\Omega_{\Lambda}\right)^{-1/2}d z\, .
\ee
Here, $\Omega_{m}$ is the fraction of the total density contributed by matter,
and $\Omega_{\Lambda}$ is the fraction contributed by the vacuum energy density.

One can get also the deceleration parameter $q_0$ and jerk parameter
$j_0$ \cite{Sahni:2002fz}:
\bea
\label{DDD1}
&& q_0 = - \left. \frac{1}{a H^2} \frac{d^2 a}{dt^2}\right|_{t=t_0}
= - \left. \frac{1}{H^2} \left\{ \frac{1}{2} \frac{d \left( H^2 \right)}{dN}
+ H^2 \right\} \right|_{N=0} \, ,\nn
&& j_0 = \left. \left\{ \frac{1}{a H^3} \frac{d^3 a}{dt^3} \right|_{t=0}
= \left. \frac{1}{2H^2} \frac{d^2 \left( H^2 \right)}{dN^2}
+ \frac{3}{2H^2} \frac{d \left( H^2 \right)}{dN} + 1 \right\} \right|_{N=0}
\, .
\eea
Here $N$ is defined by
\be
\label{e-foldings}
N \equiv - \ln \left(1+z\right)\, .
\ee
For the current time $t=t_0$, we have $N=0$. It is useful to note that
\be
\frac{d}{dN}=-(1+z)\frac{d}{dz}\, , \quad
\frac{d^{2}}{dN^{2}}=(1+z)\frac{d}{dz}+(1+z)^{2}\frac{d^{2}}{dz^{2}}\, .
\ee
Measuring the deceleration parameter and especially the jerk parameter is a
much
more difficult task than measuring $H_{0}$. In order to measure the Hubble
constant, one needs to derive the distances to objects at $\sim 100$\, Mpc;
this corresponds to a redshift of $z\gtrsim 0.02$.
To obtain $q_{0}$, one needs to observe objects to redshift
$z\gtrsim 1$.
Therefore, current observational results for the deceleration and jerk
parameters are
not totally reliable. For example, Ref.~\cite{Rapetti} gives for a flat model
tight
constraints on $q_{0}=-0.81\pm0.14$ and $j_{0}=2.16^{+0.81}_{-0.76}$ from
type Ia supernovae and X-ray cluster gas mass fraction measurements.
These results are consistent with $\Lambda$CDM
at about the 1$\sigma$ confidence level.

For the $\Lambda$CDM model, one gets
\be
q_{0}=\frac{3}{2}\Omega_{m}-1\, ,\quad j_{0}=1\, .
\ee
Therefore, measuring $j_{0}$ is important in the search for deviations from
$\Lambda$CDM, since all
$\Lambda$CDM models, regardless of the matter and cosmological constant energy
densities,
are characterized by $j=1$.

This concludes our general discussion of scalar dark energy models.
Below we present several examples of such models.

\section{Scalar Little Rip cosmology \label{SecIII}}

It is known that the Little Rip cosmology can be realized in
the class of exponential or power-law scalar potentials
\cite{Frampton-3,Astashenok}.
It is interesting that even for the case of the simplest
Klein-Gordon potential, the Little Rip can occur.

Let us start from
\be
\label{LRT}
f(\rho)=\alpha^{2}=\mathrm{const}\, .
\ee
    From the conservation law for the dark energy fluid (\ref{Fried1}), it is
easy
to
obtain
\be
\rho=\rho_{0}-3\alpha^{2}\ln(1+z).
\ee
The luminosity distance is given as
\be
D_\mathrm{L}=\frac{c}{H_{0}}(1+z)\int_{0}^{z}
\left(\Omega_{m}(1+z)^{3}+(1+3(w_{0}+1))\Omega_\mathrm{D}\right)^{-1/2}d z\, .
\ee
where the current EoS parameter $w_{0}$ is
\be
w_{0}=-1-\frac{\alpha^{2}}{\rho_{0}}\, ,
\ee
and $\Omega_{m}$ and $\Omega_\mathrm{D}$ are the matter and
dark energy contributions to the total energy budget. It is clear that if
$\alpha^{2}\ll \rho_{0}$ (and therefore $w_{0}\approx-1$) there is good
agreement with the observational data: such a cosmology is indistinguishable
from the
$\Lambda$CDM model. The deceleration and jerk parameters are given by
\be
q_{0} = \frac{3}{2}\Omega_{m}-1+\frac{3}{2}(w_{0}+1)\Omega_\mathrm{D} \, ,
\quad
j_{0}=1-\frac{9}{2}(w_{0}+1)\Omega_\mathrm{D}\, .
\ee
For $-1.15<w_{0}<-1$ and $\Omega_\mathrm{D}=0.72$, these parameters are in good
agreement with observational data. The scalar field grows with time as
\be
\phi=\phi_{0}+\alpha t\, ,
\ee
and the potential describes the massive scalar field:
\be
V(\phi)=\frac{m^{2}}{2}(\phi-\phi^{*})^{2}+\frac{\alpha^{2}}{2}\, ,\quad
m^{2}\equiv 3\alpha^{2}/2\, , \quad
\phi^{*}\equiv \phi_{0}-\frac{2\rho_{0}^{1/2}}{3^{1/2}\alpha}\, .
\ee
This type of potential is very popular in particle physics.
For example, the light scalar particles like dilaton and moduli appear in
superstring theories.
The dark energy density increases with time. Hence, the universe
accelerates. One can estimate the time required for disintegration of the
Sun-Earth system, as an example. The dimensionless inertial force
\be
\bar{F}_\mathrm{iner}=\frac{\ddot{a}}{aH_{0}^{2}}\, ,
\ee
can be expressed for $t\gg t_{0}$ as follows
\be
\bar{F}_\mathrm{iner}\approx\Omega_\mathrm{D}\frac{\rho}{\rho_{0}}.
\ee
Taking into account that the phantom energy density increases with time as
\be
\rho=\left(\rho_{0}^{1/2}-\frac{3^{1/2}}{2}(w_{0}+1)\rho_{0}t \right)^{2}\, ,
\ee
and that the Sun-Earth system disintegrates when $\bar{F}_\mathrm{iner}$ $\sim
10^{23}$ (see \cite{Frampton-2}),
one can find that the disintegration time is
\[
t_{dis}\approx\frac{10^{12}}{|w_{0}+1|H_{0}}\approx\frac{10^{13}}{|w_{0}+1|}\,
\mbox{Gyr}\, .
\]
Hence, we have presented a realistic Little Rip cosmology caused by scalar dark
energy with a standard particle physics massive potential.
Note that if such cosmology occurs, one can speculate on visible
reduction of galaxy clusters number in future cosmological surveys.

\section{Asymptotically de Sitter evolution: Pseudo-Rip \label{SecIV}}

It is evident that some phantom/quintessence models
may describe asymptotically de Sitter evolution.
However, even in this case the disintegration of bound structures may take
place. Such a scenario was dubbed the Pseudo-Rip cosmology
\cite{Frampton-2,Astashenok}.
Of course, not all asymptotically de Sitter phantom scenarios lead to
disintegration of bound structures.

Let us consider the example of a fluid which describes the phantom and/or
quintessence field asymptotically approaching the de Sitter regime. Let
\be
\label{EoSAsDS}
f(\rho)=\alpha^{2}\sin\left(\frac{\pi\rho}{\rho_\mathrm{f}}\right)\, .
\ee
If $2k\rho_\mathrm{f}<\rho<(2k+1)\rho_\mathrm{f}$, $k=0,1,...$, we have a
phantom case
while for $(2k+1)\rho_\mathrm{f}<\rho<(2k+2)\rho_\mathrm{f}$
Eq.~(\ref{EoSAsDS})
describes quintessence.
For $\rho\ll\rho_\mathrm{f}$, the parameter $w$ is nearly constant:
\bea
\label{asds}
w &\approx& -1-\frac{\alpha^{2}\pi}{\rho_\mathrm{f}}\, , \nn
\rho &=&
\rho_\mathrm{f}\left(\pm\frac{2}{\pi}\arctan\left(\tan\left(\frac{\pi\Delta}{2
(z+1)^{\delta}}\right)\right)+2k\right)\, ,\quad
\delta=3\alpha^{2}/\pi\rho_\mathrm{f}\, ,\quad
\Delta=\rho_{0}/\rho_\mathrm{f}\, ,
\eea
where the ``$\pm$'' corresponds to phantom and quintessence theories,
respectively.
Depending on the parameter $\Delta$, the dark energy density tends to a
single value from the set of ``effective cosmological constants''
\be
\Lambda^\mathrm{eff}=(2k+1)\rho_\mathrm{f}\, .
\ee
If the effective cosmological constant is sufficiently large ($\ddot{a}/a\gg
0$),
then disintegration of bound structures can occur.
It is obvious that this scenario occurs only for the phantom case (for
quintessence with asymptotic
de Sitter evolution, the acceleration of the universe can only decrease). The
dimensionless inertial force
\be
\bar{F}_\mathrm{iner}=3\frac{\ddot{a}}{\rho_{0}a}\, .
\ee
tends to 
$\rho_{f}/\rho_{0}=\Delta^{-1}$ (for $k=0$) because $a\sim\exp(\sqrt{\rho_{f}/3}t)$ at $t\rightarrow\infty$. 
Therefore if $\Delta<10^{-23}$ then 
$\bar{F}_\mathrm{iner}>10^{23}$ at $t\rightarrow\infty$ and 
the disintegration of the Sun-Earth system eventually happens.

Such a scenario is compatible with observational data.
For small $\Delta$, one can write with good accuracy the past dark energy
density
\be
\rho\approx\rho_{0}(1+z)^{-\delta}\, .
\ee
For small $\delta$ ($w\rightarrow-1$), the dark energy density is nearly
constant in the observable range $0<z<1.5$. Therefore, the observable relation
between modulus and redshift can be fulfilled in the model given
by Eq.~(\ref{EoSAsDS}).

The deceleration and jerk parameters are within the range of the standard
cosmology
(as $\delta\rightarrow 0$):
\bea
q_{0} &=& \frac{3}{2}\Omega_{m}-\frac{\delta}{2}\Omega_\mathrm{D}-1\, , \\
j_{0} &=& 1+\frac{3\delta+\delta^{2}}{2}\Omega_\mathrm{D}\, .
\eea
Thus, a realistic mild phantom scenario (Pseudo-Rip) may be easily realized.

\section{Dark energy models with a Type III future singularity \label{SecV}}

We shall consider in this section a flat universe which ends up in a Type III
future singularity \cite{Nojiri-2}. Let us start from
\be
\label{EoSBFS}
g(\rho)=-\beta^2 a_\mathrm{f}^{\epsilon}\rho^{1+\epsilon/3}\, ,
\ee
so that
\[
f(\rho)=\rho(-1+\beta^2 a_\mathrm{f}^{\epsilon}\rho^{\epsilon/3})\, ,
\]
where $\beta$, $a_\mathrm{f}$, and $\epsilon$ are positive constants.
One can find the dependence of the dark energy density on the scale factor
\be
\rho=\beta^{-6/\epsilon}\left(a_\mathrm{f}^{\epsilon}-a^{\epsilon}
\right)^{-3/\epsilon}\, .
\ee
Putting, for example, $\epsilon=1$, we have
\[
p=-\beta^2 a_\mathrm{f}\rho^{4/3}\, .
\]
In this case, one can find the scale factor in parametric form
\begin{equation}
\begin{array}{l}
a=a_\mathrm{f}\sin^{2}\eta\, ,\\
\\
\displaystyle{t=t_\mathrm{f}+\frac{1}{\kappa}\left(\ln\left|\tan\frac{\eta}{2}\right|
+\cos\eta+\frac{1}{3}\cos^{3}\eta\right)}\, ,
\end{array}
\label{solution}
\end{equation}
where
\[
\kappa=\frac{1}{2 sqrt{3} \beta^{3}}a_\mathrm{f}^{-3/2},\qquad
dt=\frac{\cos^4\eta}{\kappa\sin\eta}d\eta\, .
\]
We now set $t_\mathrm{f}=0$. Therefore $\eta=0$ corresponds to
$t=-\infty$, $\eta=\pi/2$ to $t=0$ (future singularity) and $\eta=\pi$ to
$t=+\infty$. Hence, this solution describes two universes: the
first one begins at $t=-\infty$ (Big Bang) and then
expands to a future singularity which takes place at $t=0$. The second
solution begins at $t=0$ (at a singularity) and then progressively contracts
until a big crunch singularity at $t=\infty$. The asymptotic
behavior of the scale factor near the future singularity in both cases is
\[
a=a_\mathrm{f}\left(1-(5\kappa)^{2/5}|t|^{2/5}\right),\qquad t\sim 0\, .
\]
The addition of dark matter allows us
to construct cosmological models in which the age of the universe is close to
the
conventional value of $10$-$20$\,Gyr.
The dependence of the dark energy density on redshift is given by
\begin{equation}
\label{chapl}
\rho=\rho_{0}(1+z)^{3}\left(\frac{N_{0}-1}{N_{0}(1+z)^{\epsilon}-1}
\right)^{3/\epsilon}\, ,
\end{equation}
where $N_{0}=(a_\mathrm{f}/a_{0})^{\epsilon}$. For the current value of the EoS
parameter $w_{0}$ we have
\begin{equation}
\label{w0}
w_{0}=-\beta^{2}N_{0}a_{0}^{3}\rho_{0}^{\epsilon/3}
=-\frac{N_{0}}{N_{0}-1}\, .
\end{equation}
One can use the standard relation between redshift and time
\bea
\label{age}
&& H_{0}^{-1}\int \frac{dz}{(1+z)\sqrt{h(z)}}=-\int dt\, , \nn
&&
h(z)=\Omega_{m}(1+z)^{3}+\Omega_\mathrm{D}(1+z)^{3}\left(\frac{N_{0}-1}{N_{0}
(1+ z)^{\epsilon}-1}\right)^{3/\epsilon}\, ,
\eea
for the calculation of the age of the universe and the estimation of the time
of the future singularity.
Integrating Eq.~(\ref{age}) from $z = \infty$ ($t=0$, Big Bang) to $z=0$
($t=t_{0}$) gives the age of the universe:
\begin{equation}
\label{age1}
t_{0}=H_{0}^{-1}\int^{\infty}_{0}\frac{dz}{(1+z)\sqrt{h(z)}}\, .
\end{equation}
For $N_{0}\gg 1$ (i.e., for $w_{0}\approx-1$) the function $h(z)$ can be
approximated by
\begin{equation}
\label{approx}
h(z)\approx\Omega_{m}(1+z)^{3}+\Omega_\mathrm{D}\, .
\end{equation}
Therefore, the age of the universe is eventually independent of $\epsilon$.
This parameter, however, may change the remaining time before the future
singularity $t_\mathrm{f}-t_{0}$.
Note that for the calculation of this time we can use
Eq.~(\ref{age}), simply assuming that the variable $z$ can take negative
values.
The lower limit of integration corresponds to the scale factor
$a=a_\mathrm{f}$,
i.e.,
$z_\mathrm{f}=N_{0}^{-1/\epsilon}-1$. Therefore for $t_\mathrm{f}-t_{0}$ one
gets the
following relation
\begin{equation}
\label{agef}
t_\mathrm{f}-t_{0}=H_{0}^{-1}\int^{0}_{z_\mathrm{f}}\frac{dz}{(1+z)\sqrt{h(z)}}\,
.
\end{equation}
It is obvious that our model can fit the Supernova Cosmological Project
data. For $N\gg 1$ the dark energy density is nearly constant in the
interval $0<t<t_{0}$, i.e. the model (\ref{EoSBFS}) mimics a cosmological
constant in the past but leads to a finite-time future singularity.
Moreover, the observational data do not
impose any significant restrictions on the lifetime of the universe.

The numerical calculation of the age of the universe, $t_0$, and the
difference between the future singularity time, $t_\mathrm{f}$, and $t_{0}$
for various values of $w_{0}$ and $\epsilon$ are presented
in Table~\ref{Table1}. The value of the Hubble parameter is chosen
to be $H_{0}^{-1}=13.6$\,Gyr for this calculation.

\begin{table}
\caption{A numerical calculation of the age of the universe, $t_0$, and the
difference between the future singularity time, $t_\mathrm{f}$, and $t_{0}$
for various values of $w_{0}$ and $\epsilon$. The time unit is $10^9$ years
(Gyr)
and we choose $H_{0}^{-1}=13.6$\,Gyr.
\label{Table1}}
\begin{tabular}{|c|c|c|c|c|c|c|}
\hline
    & \multicolumn{2}{c}{$w_{0}=-1.01$}\vline &
\multicolumn{2}{c}{$w_{0}=-1.05$}\vline &
\multicolumn{2}{c}{$w_{0}=-1.1$}\vline \\
\hline
$\epsilon$ & $t_\mathrm{f}-t_{0}$ & $t_{0}$ & $t_\mathrm{f}-t_{0}$ & $t_{0}$ &
$t_\mathrm{f}-t_{0}$ & $t_{0}$ \\
\hline
1 & 52.95 & 13.66 & 30.48 & 13.73 & 22.17 & 13.81 \\
2 & 29.45 & 13.66 & 17.79 & 13.71 & 13.31 & 13.78 \\
5 & 12.64 & 13.65 & 7.93 & 13.69 & 6.08 & 13.73 \\
10 & 6.40 & 13.65 & 4.09 & 13.67 & 3.17 & 13.69 \\
50 & 1.27 & - & 0.83 & - & 0.65 & - \\
100 & 0.63 & - & 0.41 & - & 0.33 & - \\
1000 & 0.06 & - & 0.04 & - & 0.03 & - \\
\hline
\end{tabular}
\end{table}

The deceleration parameter is found to be
\be
q_{0}=\frac{3}{2}\Omega_{m}-1+\frac{3}{2}\Omega_\mathrm{D}(1+w_{0})
\ee
and for $w_{0}\approx -1$ $q_{0}\approx q_{0}^\mathrm{SC}$.
For the jerk parameter we have
\be
j_{0}=\frac{3}{2}(3+\epsilon)(w_{0}+w_{0}^{2})\Omega_\mathrm{D}+1\, .
\ee
For sufficiently small $\epsilon$ and $w\approx-1$, the jerk parameter is
nearly
equal to $1$. Hence, a viable quintessence dark energy model which is
compatible
with observational data and leads to a Type III future singularity is
constructed.
Of course, it may be presented in terms of a scalar field with the field value
$\phi$ and potential $V(\phi)$ written as functions of $\eta$. Furthermore, the
dependence of $V(\phi)$ on $\phi$ can be presented explicitly.
We have
\bea
&& \phi = \frac{\eta}{\alpha_{0}}+\phi_{0}\, ,\nn
&& V(\phi) =\alpha_{1}(\cos(\alpha_{0}(\phi-\phi_{0})))^{-6/\epsilon}(1
+\cos^{-2}(\alpha_{0}(\phi-\phi_{0})))\, ,\nn
&& \alpha_{0}=\sqrt{\frac{1}{12}}\epsilon\, ,\qquad
\alpha_{1}=\frac{1}{2}\beta^{-6/\epsilon}a_\mathrm{f}^{-3}\, .
\eea
Other models with similar properties can be constructed.

\section{Type II future singularity dark energy \label{SecVI}}

In this section we discuss realistic models of dark energy
which contain a Type II future singularity \cite{Nojiri-2} (or sudden future
singularity \cite{Barrow}). It is known that such evolution may be also
realized in $f(R)$ modified gravity \cite{Bamba}.

The simplest choice for an EoS producing a Type II future singularity is
\be
\label{TM}
f(\rho)=\frac{\alpha^{2}}{1-\rho/\rho_\mathrm{f}}\, ,
\ee
where $\alpha$ and $\rho_\mathrm{f}$ are positive constants.
For $\rho_{0}<\rho_\mathrm{f}$, such a model describes phantom energy.
Its energy density grows with time until
the pressure tends to infinity and a phantom sudden future singularity occurs.
If $\rho_{0}>\rho_\mathrm{f}$, the energy density decreases and a big crush
occurs when $\rho$ is equal to $\rho_\mathrm{f}$.

\begin{figure}
\rotatebox{90}{\includegraphics[scale=0.5,angle=270]{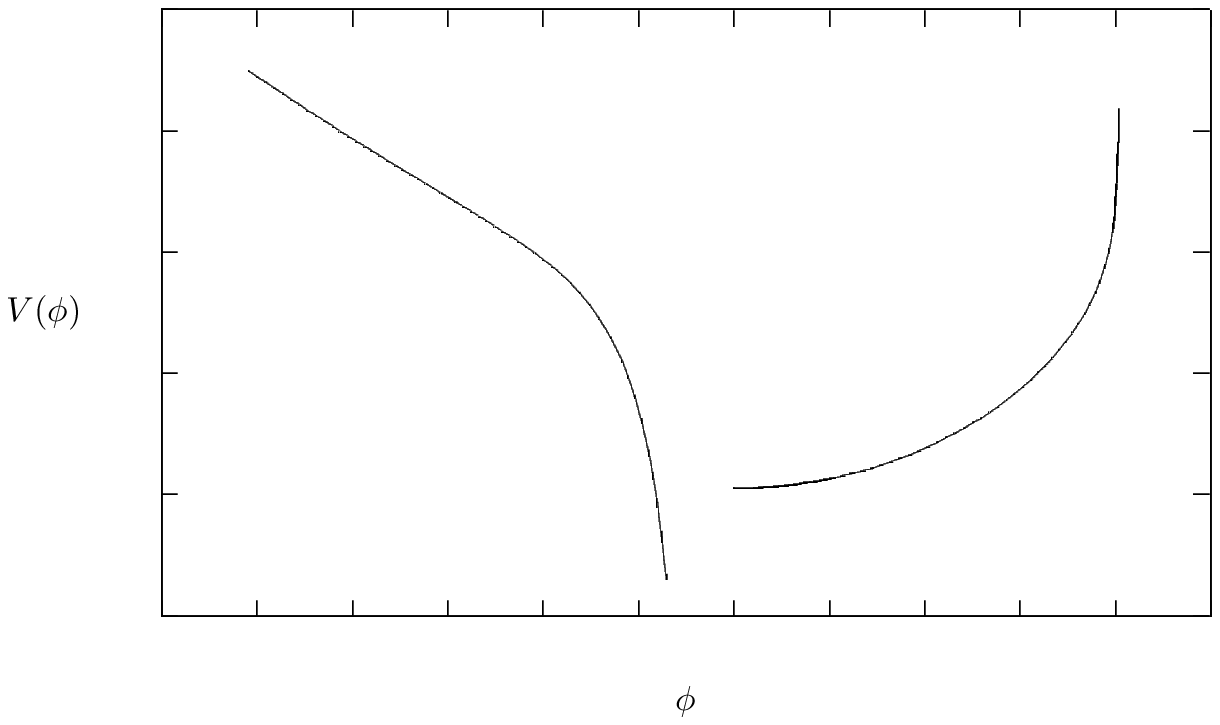}}\\
\caption{The scalar potential for the model (\ref{TM}). For a crush the
scalar field rolls down from $-\infty$ to $0$ and
$V(\phi)\rightarrow\-\infty$; for a phantom sudden future singularity the
scalar
field
rolls up and $V(\phi)\rightarrow\infty$ at some $\phi=\phi_{s}$.}\label{1}
\end{figure}

The time remaining before the future singularity is
\be
t_\mathrm{f}-t_{0}=\frac{2}{\sqrt{3}}\int_{x_{0}}^{x_\mathrm{f}}\frac{dx}{\alpha^2}
\left(1 -\left(\frac{x}{x_\mathrm{f}}\right)^{2}\right)\, .
\ee
The corresponding description in terms of scalar field theory can be derived
through Eqs.~(\ref{Vx}) and (\ref{phix}). The potential of the scalar field in
parametric form is:
(i) for phantom energy
\bea
\phi(y)&=&\frac{1}{\sqrt{3\gamma}}\left(\arcsin y+y(1-y^{2})^{1/2}\right)\, ,\\
V(y)& =& \frac{\rho_\mathrm{f}}{2}\left(2y^{2}+\frac{\gamma}{1-y^{2}}\right)\,
,\quad
0\leq y\leq 1\, ,
\eea
(ii) for quintessence
\bea
\phi(y)&=&-\frac{1}{\sqrt{3\gamma}}\left(\frac{\sqrt{1-y^{2}}}{y^{2}}-\ln(1-
\sqrt{1-y^{2}})+\ln y\right)\, ,\\
V(y)& =&
\frac{\rho_\mathrm{f}}{2}\left(2y^{-2}+\gamma\frac{y^{2}}{y^{2}-1}\right)\,
,\quad
0\leq y\leq 1\, ,
\eea
and $\gamma=\alpha^{2}/\rho_\mathrm{f}$. This potential is depicted in
Figure~\ref{1}.

Such a model in principle can fit the latest
supernova data from the Supernova Cosmology Project.
The dependence of the dark energy density on the redshift $z$ can be derived
from
Eq.~(\ref{arho}). After a simple algebraic calculation, one can obtain
\be
\rho=\rho_\mathrm{f}\left(1\pm\left(\left(1-\Delta\right)^2
+6\gamma\ln(1+z)\right)^{1/2}\right)\, ,\quad
\Delta=\rho_{0}/\rho_\mathrm{f},\quad
\gamma=\alpha^{2}/x_\mathrm{f}^{2}\, .
\ee
The sign ``$+$'' corresponds to the case of quintessence ($\Delta>1$) while
sign
``$-$'' to that of phantom energy ($\Delta<1$).
The current EoS parameter $w_{0}$ is
\be
w_{0}=-1-\frac{\gamma}{\Delta(1-\Delta)}\, .
\ee
Therefore, the dependence of the luminosity distance $D_\mathrm{L}$ on the
redshift
$z$
is
\bea \label{DL}
D_\mathrm{L} &=& \frac{c}{H_{0}}(1+z)\int_{0}^{z}\left(\Omega_{m}
(1+z)^{3}+\Omega_\mathrm{D}h(z)\right)^{-1/2}d z\, ,\nn
h(z) &=&
\Delta^{-1}\left(1\pm\left(\left(1-\Delta\right)^2+6\gamma\ln(1+z)\right)^{1
/2}\right)\, .
\eea
Eq.~(\ref{DL}) coincides with (\ref{DLSC}) if $\gamma=0$ ($f(x)=0$). The
SNe data are available in the range $0<z<1.5$. Therefore, if the parameters
$\Delta$ and $\gamma$ are such that $(1-\Delta)^{2}\ll 6\gamma\ln(1+z)$ in the
observable range, the model under discussion is indistinguishable from
$\Lambda$CDM cosmology.

The time remaining before a future singularity is
\be
t_\mathrm{f}-t_{0}=\frac{1}{H_{0}}\int_{u}^{0}du(1+u)^{-1}
\left(\Omega_{m}(1+u)^{3}+\Omega_\mathrm{D}h(u)\right)^{-1/2}\, .
\ee
The variable $u=a_{0}/a-1$ varies from $0$ (present time) to
$\exp(-(1-\Delta)^{2}/6\gamma)-1$ ($p\rightarrow\infty$).
The function $h(u)$ coincides with $h(z)$ in
Eq.~(\ref{DL}) (after changing $z\rightarrow u$).
Numerical estimation of the difference $t_\mathrm{f}-t_{0}$ for various values
of $\Delta$ and $\gamma$ is given in Table~\ref{Table II} and
Table~\ref{Table III}.
For quintessence, the difference
$t_\mathrm{d}-t_{0}$ ($t_\mathrm{d}$ is the moment of time when $\ddot{a}=0$
and
deceleration begins) is also calculated.
We use the value of the Hubble parameter $H^{-1}_{0}=13.6$\,Gyr.

\begin{table}
\caption{Numerical estimation of the difference $t_\mathrm{f}-t_{0}$
in Gyr for various values
of $\Delta$ and $\gamma$ in the case of a phantom model. \label{Table II}}
\begin{tabular}{|c|c|c|c|c|c|}
\hline
$\Delta$ & $w_{0}=-1.10$ & $w_{0}=-1.08$ & $w_{0}=-1.06$ & $w_{0}=-1.04$ &
$w_{0}=-1.02$ \\
\hline
0.5 & 22.6 & 28.4 & 38.2 & 57.6 & 116.1 \\
0.75 & 7.3 & 9.9 & 12.5 & 20.3 & 39.8 \\
0.95 & 1.1 & 1.5 & 2.0 & 3.1 & 6.3 \\
\hline
\end{tabular}
\end{table}

\begin{table}
\caption{Numerical estimation of the difference $t_\mathrm{f}-t_{0}$
in Gyr for various values
of $\Delta$ and $\gamma$ in the case of quintessence. \label{Table III}}
\begin{center}
\begin{tabular}{|c|c|c|c|c|c|c|c|c|c|c|c|c|}
\hline
& \multicolumn{2}{c}{$w_{0}=-0.98$}\vline &
\multicolumn{2}{c}{$w_{0}=-0.96$}\vline & \multicolumn{2}{c}{$w_{0}=-0.94$}
\vline & \multicolumn{2}{c}{$w_{0}=-0.92$} \vline &
\multicolumn{2}{c}{$w_{0}=-0.90$} \vline \\
\hline
$\Delta$ & $\Delta t_\mathrm{d}$ & $\Delta t_\mathrm{f}$ & $\Delta
t_\mathrm{d}$ & $\Delta
t_\mathrm{f}$ & $\Delta t_\mathrm{d}$ & $\Delta t_\mathrm{f}$ & $\Delta
t_\mathrm{d}$ & $\Delta t_\mathrm{f}$ &
$\Delta t_\mathrm{d}$ & $\Delta t_\mathrm{f}$ \\
\hline
1.05 & 5.78 & 5.79 & 2.8 & 2.81 & 1.84& 1.86 & 1.36 & 1.38 & 1.08 & 1.10
\\
1.25 & 26.70 & 26.74 & 12.88& 12.96 & 8.34 & 8.45 & 6.10 & 6.24 & 4.76&
4.94 \\
1.5 & 46.32 & 46.42 & 22.52 & 22.72 & 14.57 & 14.85 & 10.60 & 10.96 &
8.23 & 8.65 \\
\hline
\end{tabular}\end{center}
\end{table}

The difference $\delta \mu=5\log (D/D^\mathrm{SC})$ ($\mu$ is the distance
modulus)
for
$0<z<1.5$ lies in the interval ($-0.35$ to $0.35$) for these parameter values.
Taking into account
that errors in the definition of the SNe modulus are $\sim 0.075\div 0.5$, we
conclude
that our model fits these data with excellent precision. Note that futher
observational support for quintessence which leads to a Type II future
singularity is given in Ref.~\cite{Dab}.

The deceleration and jerk parameters are given by
\bea
\label{DDD1B}
q_0 &=& \frac{9\gamma}{2\Delta(1-\Delta)}\Omega_\mathrm{D}
+ \frac{3}{2} \Omega_m - 1\, ,\nn
j_0 &=& -\frac{9\gamma}{2\Delta(1-\Delta)}\left(1
+\frac{\gamma}{(1-\Delta)^{2}}\right)\Omega_\mathrm{D} + 1\, .
\eea
It is convenient to present $q_{0}$ and $j_{0}$ through the parameters $w_{0}$
and $\Delta$:
\bea
q_0 &=& -\frac{9}{2}(w_{0}+1)\Omega_\mathrm{D}+\frac{3}{2} \Omega_m - 1\, ,\nn
j_0 &=& \frac{9}{2}(w_{0}+1)\left(1-\frac{\Delta}{1-\Delta}(w_{0}+1)\right)
\Omega_\mathrm{D} + 1\, .
\eea
For $w_{0}\approx -1$, the jerk parameter $j_0$ differs significantly from
the standard value only for $\Delta\rightarrow 1$. The deceleration parameter
$q_{0}$ for $-1.05<w<-0.95$ lies in the interval
$q_{0}^\mathrm{SC}-0.16<q_{0}<q_{0}^\mathrm{SC}+0.16$. Taking into account
the errors in the definition of $q_{0}$ and $j_{0}$ one can conclude that
dark energy models with a possible big crush or sudden future singularity fit
well the current observational data.

The example considered above is a good theoretical illustration of dark
energy models mimicking vacuum energy but leading to singularities of Type II.
Dark energy with such behavior can
be realized if the function $f(x)$ has a singularity at $x=x_\mathrm{f}$.

An important remark is in order. For quintessence
the disintegration of bound structures before a future singularity seems to be
impossible.  From Eqs.~(\ref{EoS}) and (\ref{Fin}) it follows that
\be
F_\mathrm{iner}=-ml\frac{1}{2}\left(w+\frac{1}{3}\right)\rho\, ,
\ee
Therefore, the maximal value of the inertial force for quintessence is
\[
F_\mathrm{iner}^\mathrm{max}=\frac{ml\rho}{3}\, .
\]
The energy density of quintessence decreases when the universe expands and
the inertial force also decreases with time. If we consider an EoS for which
$f(\rho)$ changes sign at $x=x_\mathrm{ph}$ then the energy density increases
with
time and tends to $x_\mathrm{ph}^{2}$. Hence, such an expanding universe is a
de Sitter one.

\section{Big crush dark energy models \label{SecVII}}

Another way to construct cosmological models with various
types of evolution is
to define the Hubble parameter as a function of time.
Let us consider the Type II singularity, for example.
This case occurs when $H$ is finite and $\dot H$ diverges but is
negative. In this case,
even though the universe is expanding, all structures are crushed rather than
ripped. An example is given by
\be
\label{lc1}
H = H_0^{(0)} + H_1^{(0)} \left(t_c - t\right)^\alpha\, .
\ee
Here $H_0^{(0)}$ and $H_1^{(0)}$ are positive constants and $\alpha$ is a
constant with $0<\alpha<1$.
If we choose $\alpha$ to be given by the inverse of an odd number,
$\alpha=1/(2n+1)$,
with positive integer $n$, we can extend $H$ beyond $t=t_c$ by defining
\be
\label{H1}
H = \left\{
\begin{array}{ll}
H_0^{(0)} + H_1^{(0)} \left(t_c - t\right)^\alpha & \mbox{when}\ t<t_c \\
H_0^{(0)} - H_1^{(0)} \left(t - t_c\right)^\alpha & \mbox{when}\ t>t_c
\end{array} \right. \, .
\ee
In a sense, the $H$ obtained here is a smooth function of $t$ although $\dot H$
diverges at $t=t_c$ since there is no sharp point, that is, a point where the
line folds.
At the point $\dot H$ diverges but $H$ is the continuous, single valued,
and monotonically increasing or decreasing function of $t$.
One may consider the following model
\be
\label{H2}
H = \left\{
\begin{array}{ll}
H_0^{(0)} + H_1^{(0)} \left(\tanh\frac{\left(t_c - t\right)}{t_0}\right)^\alpha
& \mbox{when}\ t<t_c \\
H_0^{(0)} - H_1^{(0)} \left(\tanh\frac{\left(t - t_c\right)}{t_0}\right)^\alpha
& \mbox{when}\ t>t_c
\end{array} \right. \, .
\ee
In the model (\ref{H2}), we find $H\rightarrow\mbox{const}$ when
$t\to\pm\infty$,
that is, the space-time is asymptotically de Sitter.
Note that when $H$ is finite, a rip occurs only for the phantom case, since
$\dot H>0$
before the
singularity. However, a crush occurs for quintessence with $\dot H<0$.
In the models (\ref{H1}) and (\ref{H2}), a rip occurs when $H_1^{(0)}<0$
and a crush occurs when $H_1^{(0)}>0$.

 From the FRW equations (\ref{Fried1}), the total energy density and pressure
of dark energy as functions of time are:
\be
\rho(t)+\rho_{m}(t)=3H^{2}\, ,\quad
p(t)=-3H^{2}-2\dot{H}\, .
\ee
Neglecting the matter density, one can easily obtain the EoS of dark energy
corresponding to model (\ref{H1}):
\be
p=\left\{
\begin{array}{ll}
-\rho+2\alpha 3^{n} H_{1}^{(0)}(\rho^{1/2}-\rho_\mathrm{f}^{1/2})^{-2n}\, ,
\quad \rho_\mathrm{f}=3H_{0}^{(0)2} & \mbox{when}\ t<t_c \\
-\rho-2\alpha 3^{n} H_{1}^{(0)}(\rho^{1/2}-\rho_\mathrm{f}^{1/2})^{-2n}\, ,
\quad \rho_\mathrm{f}=3H_{0}^{(0)2} & \mbox{when}\ t>t_c
\end{array} \right. \, .
\ee
When $H_{1}^{(0)}<0$ or $H_{1}^{(0)}>0$, $p\rightarrow-\infty$ (phantom Type II
singularity) or
$p\rightarrow+\infty$ (quintessence Type II singularity) at $t\rightarrow
t_{c}-0$.
The scale factor is finite at $t\rightarrow t-0$.
\be
a(t)=a_{0}\exp\left\{H_{0}^{(0)}t+H_{1}^{(0)}(t_{c}-t)^{\alpha+1}/(\alpha+1)
    -H_{1}^{(0)}t_{c}^{\alpha+1}/(\alpha+1)\right\}\, .
\ee

In the following, we assume $\alpha$ is given by $\alpha=1/(2n+1)$
with positive integer $n$.
We now investigate if any object can be ripped or crushed at the singularity
$t=t_c$ although
the inertial force $F_\mathrm{iner}$ diverges. For this purpose, we consider
the work or the shift of
the kinetic energy of the particle.  For purposes of
this estimation, we neglect all forces aside from the inertial force, and we
neglect the first term
in (\ref{i1}) since we assume only $\dot H$ diverges. Then by solving the
equation of motion
\be
\label{H3}
m \ddot x = F_\mathrm{iner}\sim m l \dot H
= - m l H_1^{(0)} \alpha\left(t_c - t\right)^{\alpha - 1}
\, ,
\ee
for the model (\ref{H1}), one finds
\be
\label{H4}
x = x_0 + v_0 t - \frac{l H_1^{(0)}}{\alpha+1}\left(t_c - t\right)^{\alpha+1}\,
.
\ee
Then the shift of the kinetic energy can be estimated to be
\be
\label{H5}
\Delta T = \int F \dot x dt \sim
  - m v_0 H_1^{(0)} \alpha \int \left(t_c - t\right)^{\alpha-1} dt
+ m l^2 {H_1^{(0)}}^2 \alpha \int \left(t_c - t\right)^{2\alpha - 1} dt \, .
\ee
Since $\alpha>0$, the integration and therefore $\Delta T$ is finite.
Hence if the magnitude of the binding energy or the energy supporting the
bound object is larger than the absolute value of $\Delta T$, the object
is not ripped or crushed although $F_\mathrm{iner}$ is infinite at $t=t_c$.

In the case of a Type III singularity, $H$ behaves as in (\ref{lc1}) but
$\alpha$ is negative and greater than unity, $-1<\alpha<0$.
We also note that $H_1^{(0)}>0$ since $H>0$ when $t<t_c$.
Then in the inertial force (\ref{i1}), the first term behaves as
$\dot H\sim \left(t_c - t\right)^{\alpha-1}$ and the second one as
$H^2 \sim \left(t_c - t\right)^{2\alpha}$. Since $-1<\alpha<0$, the
first term dominates. Then in a way similar to (\ref{H5}),
the shift of the kinetic energy can
be estimated as
\be
\label{H5b}
\Delta T \sim - m v_0 H_1^{(0)} \left(t_c - t\right)^\alpha
+ \frac{m l^2 {H_1^{(0)}}^2}{2} \left(t_c - t\right)^{2\alpha} \, ,
\ee
when $t<t_c$ and it becomes positive and diverges when $t\to t_c$.
Therefore the  Rip surely occurs even for a Type III singularity,
which is different from the Type II singularity in (\ref{H5}), where
all the objects are not always crushed or ripped.

By using the formulation in Ref.~\cite{Nojiri:2005pu},
we now consider what kind of scalar tensor model, whose action is given by
\be
\label{ma7}
S=\int d^4 x \sqrt{-g}\left\{
\frac{1}{2\kappa^2}R - \frac{1}{2}\omega(\phi)\partial_\mu \phi
\partial^\mu\phi - V(\phi) \right\}\, ,
\ee
can realize the evolution of $H$ given by Eq.~(\ref{lc1}).
Here, $\omega(\phi)$ and $V(\phi)$ are functions of the scalar field $\phi$.
If we consider the model where $\omega(\phi)$ and $V(\phi)$ are given by a
single function $f(\phi)$, as follows,
\be
\label{ma10}
\omega(\phi)=- \frac{2}{\kappa^2}f''(\phi)\, ,\quad
V(\phi)=\frac{1}{\kappa^2}\left(3f'(\phi)^2 + f''(\phi)\right)\, ,
\ee
the exact solution of the FRW equations has the following form:
\be
\label{ma11}
\phi=t\, ,\quad H=f'(t)\, .
\ee
Then for the model (\ref{lc1}) with $\alpha=1/(2n+1)$, we find
\be
\label{H6}
\omega(\phi) = \frac{2H^{(0)}_1 \alpha}{\kappa^2}\left( t_c - \phi
\right)^{- \frac{2n}{2n+1}}\, ,\quad
V(\phi) = \frac{1}{\kappa^2} \left\{ \left( H^{(0)}_0 + H^{(0)}_1
\left( t_c - \phi \right)^{- \frac{1}{2n+1}} \right)^2
   - 2H^{(0)}_1 \alpha\left( t_c - \phi \right)^{- \frac{2n}{2n+1}} \right\}\, .
\ee
If we redefine the scalar field as
\be
\label{H7}
\varphi = - \frac{\sqrt{(2n+1) H^{(0)}_1}}{\kappa (n+1)}
\left( t_c - \phi \right)^{\frac{n+1}{2n+1}}\, ,
\ee
the kinetic term in the action (\ref{ma7}) becomes canonical
\be
\label{H8}
    - \frac{1}{2}\omega(\phi) \partial_\mu \phi \partial^\mu\phi
= - \frac{1}{2} \partial_\mu \varphi \partial^\mu\varphi \, ,
\ee
and the potential is given by
\be
\label{H9}
V(\phi) = \frac{1}{\kappa^2} \left\{ \left( H^{(0)}_0 + H^{(0)}_1
\left( - \frac{\kappa (n+1)}{\sqrt{2H^{(0)}_1 (2n+1)}}\varphi
\right)^{\frac{1}{n+1}} \right)^2
    - H^{(0)}_1 \alpha \left( - \frac{\kappa (n+1)}{\sqrt{2H^{(0)}_1 (2n+1)}}
\varphi\right)^{-\frac{2n}{n+1}} \right\}\, .
\ee
Note that $\varphi<0$ when $\phi=t<t_c$ and $\varphi\to 0$ when $\phi=t\to
t_c$.
Near the singularity $\phi=t\to t_c$ ($\varphi\to 0$), only the last term
in the potential (\ref{H9}) dominates:
\be
\label{H10}
V(\phi) \sim - \frac{H^{(0)}_1 \alpha }{\kappa^2} \left( - \frac{\kappa
(n+1)}{\sqrt{2H^{(0)}_1 (2n+1)}}
\varphi\right)^{-\frac{2n}{n+1}} \, .
\ee
In particular, when $n=1$, we find
\be
\label{H11}
V(\phi) \sim \frac{\alpha}{\kappa^3} \sqrt{\frac{3 {H^{(0)}_1}^3}{2}}
\varphi^{-1}\, .
\ee
Thus, the big crush occurs when the scalar field drops into the infinitely
deep potential proportional to the inverse power of the scalar potential.

We have constructed models which generate the big crush and have given the
explicit action in terms of the scalar field. After the big crush, the universe
may evolve
to asymptotic de Sitter space-time.
Hence, big crush phenomenon looks much less dangerous than disintegration of
bound structures.

\section{Phantom models and singularities \label{SecPhantom}}

One way to realize many of the models proposed here is through a
scalar field with a negative kinetic term (phantom models).  The asymptotic
future evolution of such models was examined systematically in Ref.~\cite{KSS},
and we restate a number of those results here in order to show explicitly the
relation between various types of future singularity.

The simplest phantom models are characterized by a field $\phi$ with a negative
kinetic term.  Such models evolve
according to the equation
\begin{equation}
\label{phiev}
\ddot{\phi}  +   3 H \dot{\phi} - V^\prime(\phi) = 0\, ,
\end{equation}
where the prime denotes the derivative with respect
to $\phi$.  A field evolving according to this equation rolls uphill in the
potential.  The density and pressure for the phantom field are given by
\be
\rho_{\phi}=-\frac{1}{2}\dot \phi^2 + V(\phi)\, ,
\ee
and
\be
p_{\phi}=-\frac{1}{2}\dot \phi^2 - V(\phi)\, ,
\ee
respectively, so the equation of state parameter is
\begin{equation}
\label{w}
w_{\phi} =\frac{(1/2)\dot \phi^2 + V(\phi)}{ (1/2)\dot \phi^2 - V(\phi)}\, .
\end{equation}

As noted in Ref.~\cite{KSS}, the asymptotic behavior of the equation of state
parameter depends on the corresponding asymptotic behavior of $V^\prime/V$.
If $V^\prime/V \rightarrow 0$, then $w \rightarrow -1$.  This set of models
displays the most diverse behavior, since it can correspond to either
a Big Rip, a Little Rip, or a Pseudo-Rip, depending on the exact functional
form for $V(\phi)$.  A Big Rip (Type I singularity) occurs when \cite{KSS}
\be
\label{ripcondition}
\int \frac{\sqrt{V(\phi)}}{V^\prime(\phi)} d\phi \rightarrow \mbox{finite.}
\ee
If, instead, the integral in equation (\ref{ripcondition}) diverges, we have
either a Little Rip or a Pseudo-Rip. A Pseudo-Rip occurs if $V(\phi)
\rightarrow \mbox{const}$, while a Little Rip occurs if $V(\phi) \rightarrow
\infty$ (see also Section~\ref{SecII}).

The second set of models examined in Ref.~\cite{KSS} corresponds
to $V^\prime/V \rightarrow constant$.  This gives a constant value
for $w$ with $w < -1$, and produces a Big Rip (Type I) singularity.

Finally, if $V^\prime/V \rightarrow \pm \infty$, we have
$w \rightarrow - \infty$, which can result in a Type III singularity
(see also Ref.~\cite{Sami:2003xv}).

\section{Conclusion \label{SecVIII}}

In summary, dark energy models with various scenarios of evolution have been
presented.
Specifically, we constructed scalar dark energy models with Type II and Type
III finite-time future singularities, Little Rip and Pseudo-Rip cosmologies with
finite-time disintegration of bound structures and Big Crush cosmologies. It was shown
that such models are consistent with observational data from the
Supernova Cosmology Project and therefore may be viable alternatives to the
$\Lambda$CDM
cosmology. Moreover, they may be stable for billions of years before
entering
a soft future singularity (with a finite scale factor at the Rip) or before
entering a finite-time dissolution of bound structures.

We have shown that the future evolution of the universe is
determined
by the selected EoS of dark energy. Unfortunately, current data for such
important parameters as $q_{0}$ and $j_{0}$ are not very reliable, so the
nature of the dark energy
cannot yet be determined, and one can therefore
only consider some typical models.
In the future, more accurate measurements of the
deceleration and jerk parameters as well as other cosmological parameters will
help to define the exact nature of dark energy.
Then we will acquire the information on the parameters
of the fluid description for the EoS of dark energy in this paper
and therefore also the information of the parameters in
a reconstructed scalar field theory. 
In other words, more precise values of cosmological parameters may 
significantly constrain the dark energy models under discussion.

The key point is that the current observational data do not answer, even in
principle,
the question of whether or not the universe will end in a future
singularity or Rip cosmology. One can construct (as we have here) models that
mimic standard
$\Lambda$CDM up the present, but evolve in the future into any number of
possible
future states, including Pseudo-Rip models, Little Rip models, and a variety of
different
future singularities. With a variety of $\Lambda$CDM-like cosmological models
in hand, one can already start to think about future cosmological
experiments to define the future of the universe more precisely.

\section*{Acknowledgments}

S.N. is supported by Global COE Program of Nagoya University (G07)
provided by the Ministry of Education, Culture, Sports, Science \&
Technology and by the JSPS Grant-in-Aid for Scientific Research (S) \# 22224003
and (C) \# 23540296.
The work by SDO has been supported in part by MICINN (Spain) project
FIS2010-15640,
by AGAUR 2009SGR-994 and by JSPS Visitor Program S11135 (Japan).
R.J.S. is supported in part by the Department of Energy (DE-FG05-85ER40226).

\end{document}